\documentclass[runningheads]{llncs}
\usepackage{graphicx}
\usepackage{xcolor}
\usepackage[export]{adjustbox}
\usepackage{graphics}
\usepackage{caption}
\usepackage{subfig}
\usepackage{cite}

\usepackage{booktabs}
\usepackage{multirow}
\usepackage{tabularx}

\usepackage[margin=1in]{geometry}
\usepackage{array}
\newcolumntype{L}[1]{>{\raggedright\let\newline\\\arraybackslash\hspace{0pt}}m{#1}}
\newcolumntype{C}[1]{>{\centering\let\newline\\\arraybackslash\hspace{0pt}}m{#1}}
\newcolumntype{R}[1]{>{\raggedleft\let\newline\\\arraybackslash\hspace{0pt}}m{#1}}

\begin{document}
\title{Cough Detection Using Hidden Markov Models}
%
%
\author{Aydin Teyhouee\inst{1}\orcidID{0000-0001-8822-9413} \and
Nathaniel D. Osgood\inst{3}\orcidID{0000-0002-3994-3298}}
\authorrunning{A. Teyhouee et al.}
%
\institute{University of Saskatchewan, Saskatchewan, Canada\\
\email{aydin.teyhouee@usask.ca}\\
\email{osgood@cs.usask.ca}
}
\maketitle              
\begin{abstract}
Respiratory infections and chronic respiratory diseases impose a heavy health burden worldwide. Coughing is one of the most common symptoms of many such infections, and can be indicative of flare-ups of chronic respiratory diseases.  Whether at a clinical or public health level, the capacity to identify bouts of coughing can aid understanding of population and individual health status.  Developing health monitoring models in the context of respiratory diseases and also seasonal diseases with symptoms such as cough has the potential to improve quality of life, help clinicians and public health authorities with their decisions and decrease the cost of health services. In this paper, we investigated the ability to which a simple machine learning approach in the form of Hidden Markov Models (HMMs) could be used to classify different states of coughing using univariate (with a single energy band as the input feature) and multivariate (with a multiple energy band as the input features) binned time series using both of cough data. We further used the model to distinguish cough events from other events and environmental noise.  Our Hidden Markov algorithm achieved 92\% AUR (Area Under Receiver Operating Characteristic Curve) in classifying coughing events in noisy environments. Moreover, comparison of univariate with multivariate HMMs suggest a high accuracy of multivariate HMMs for cough event classifications.

\keywords{Cough detection  \and Machine learning \and Hidden Markov Model \and Frequency domain \and Time domain \and Pattern recognition \and Spectrogram \and Health care \and Public health.}
\end{abstract}
\section{Introduction}
Symptoms such as cough are important clinical signs. Coughing is the most common symptom in respiratory diseases, and awareness of the occurrence or persistent presence of a cough can provide valuable information to physicians. Detailed awareness of coughing can aid physicians with their treatment on the basis of quantitative assessments such as frequency or intensity as well as qualitative assessments such as dry or wet coughs \cite{Swarnkar}. Moreover, cough detection analysis has the potential to reduce the cost of health services by -- for example -- detecting the early signs of diseases and making preemptive diagnosis possible and prescribing basic treatments while they are still effective \cite{Larson}. However, the benefits of securing reliable, and timely quantification of coughing behavior can also offer benefits beyond the physician's office.  Collecting cough data using monitoring devices such as mobile sensors or other devices and analyzing the audio signals of coughs can support remote monitoring of patients with chronic respiratory illnesses or restricted mobility. For such diseases, awareness of flare-ups of coughing can motivate the need to present for care, and can inspire changes to treatment recommendations.  A final and important advantage of cough recognition resides in its potential to provide health authorities with timely surveillance information about emergence of high-burden respiratory conditions, thereby supporting earlier outbreak identification in particular geographic areas, thereby better supporting public health decision making, including the design of public health interventions. 

The duration of a cough sound typically varies between 0.2 and 1 second \cite{Korpas}, and exhibits a sequence of distinct acoustic patterns. The origin of these patterns is airway narrowing and bifurcation. The airway narrowing is due to a change in the thickness of the airflow walls (inflammation, mucus collection, bronchoconstriction and fibrosis). A typical cough sound usually is composed of three stages: an explosive expiration due to the abrupt opening of glottis, the intermediate stage in which cough sounds are reduced, and the voiced stage due to the closing of the vocal cord. There are a variety of patterns of coughing based on the presence or absence of each of these stages \cite{Morice}.

\begin{figure}
\includegraphics[width=\linewidth]{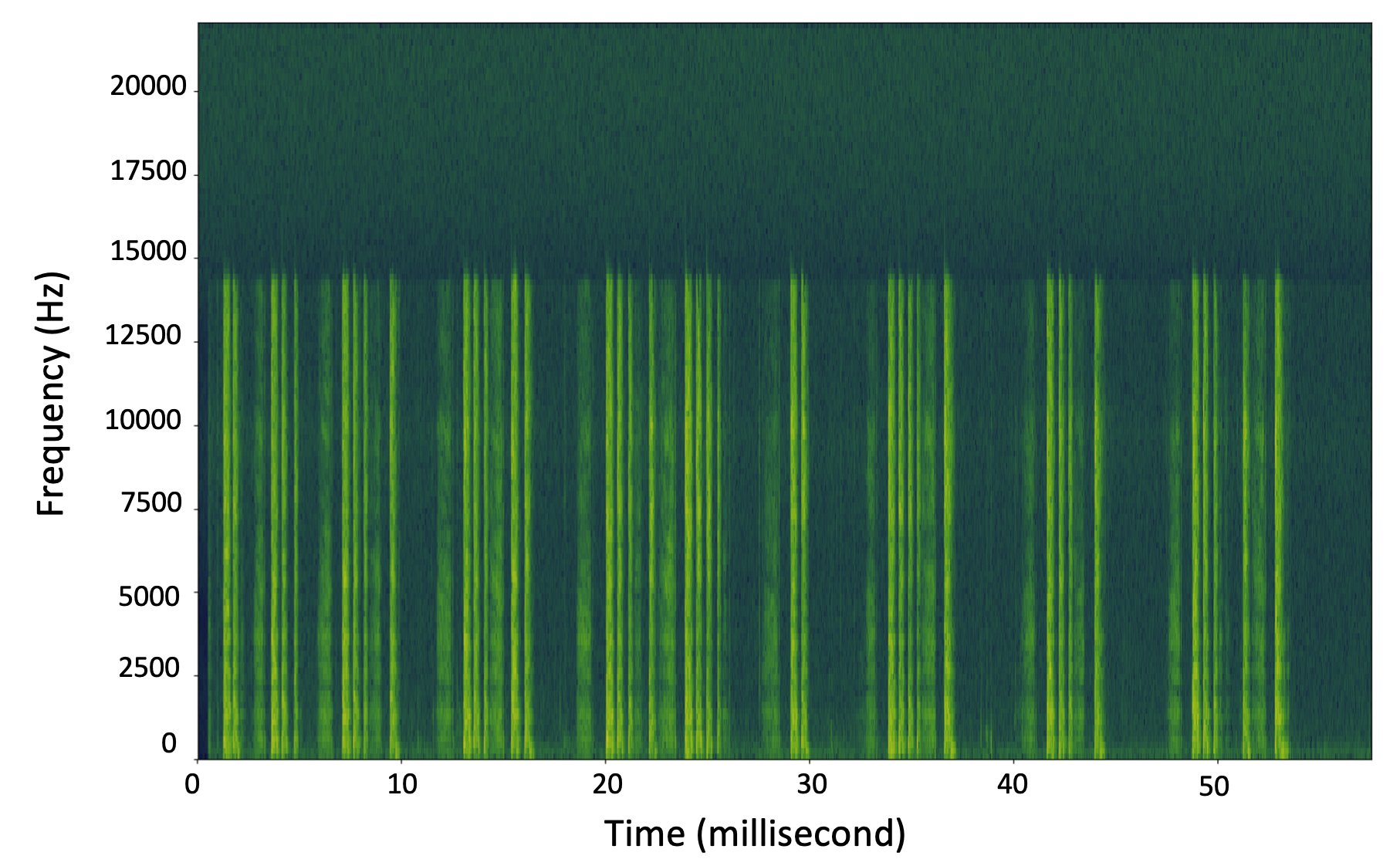}
\caption{A spectrogram of a sample cough} \label{fig:spectrogram}
\end{figure}

A visual representation of the spectrum of frequencies of a cough signal as it varies over time is shown in the spectrogram of figure \ref{fig:spectrogram}, which is depicted as a heat map, with the lowest and highest intensities being represented by dark and light green, respectively.

Several studies have described methods to analyze cough characteristics, considering the subjective interpretation of cough sound recordings and the analysis of spectrograms \cite{Korpas2, Day, Doherty, Murata, Thorpe, Toop}. There are two main research streams for cough recognition. One stream investigates audio signals frame-by-frame and combines consecutive cough frames as a cough event \cite{frame}. The second stream consists of event detection and cough classification steps. Event detection identifies cough event candidates; each candidate is then classified as a cough or non-cough event \cite{event}. Our work follows the first stream, by seeking to detect cough signals in continuous audio recording using a Hidden Markov Model (HMM).


This paper investigated the performance of an HMM, where each state of that model corresponds to a portion of a typical cough, and where observables represent summaries of information from sound profiles. We further investigated the performance of the model in detecting each state and thus distinguishing a period of time in which a cough was occurring from when it was not.  The HMM could further be used to distinguish coughing from non--coughing behaviour when considering a longer period of time, and when the main focus is to identify bouts of cough present in an sound recording events. To achieve this, the acoustic energy was selected as the observable and measurable feature to feed into a univariate HMM. In another attempt, using the frequency or pitch of the sound, the energy spectrum as the observation input was split into a vector of three sub-features as low, mid and high energy bands. Finally, we compared the performance of these two scenarios. 

\section{Materials and Methods}
\subsection{Data Collection and Labeling}
The cough data used in this article is collected from recordings of cough sounds from individuals in Computational Epidemiology and Public Health Informatics Laboratory in the Department of Computer Science at the University of Saskatchewan.  A duration of 20 minutes of such cough sounds were manually annotated by the authors.

\begin{figure}
\includegraphics[width=\linewidth]{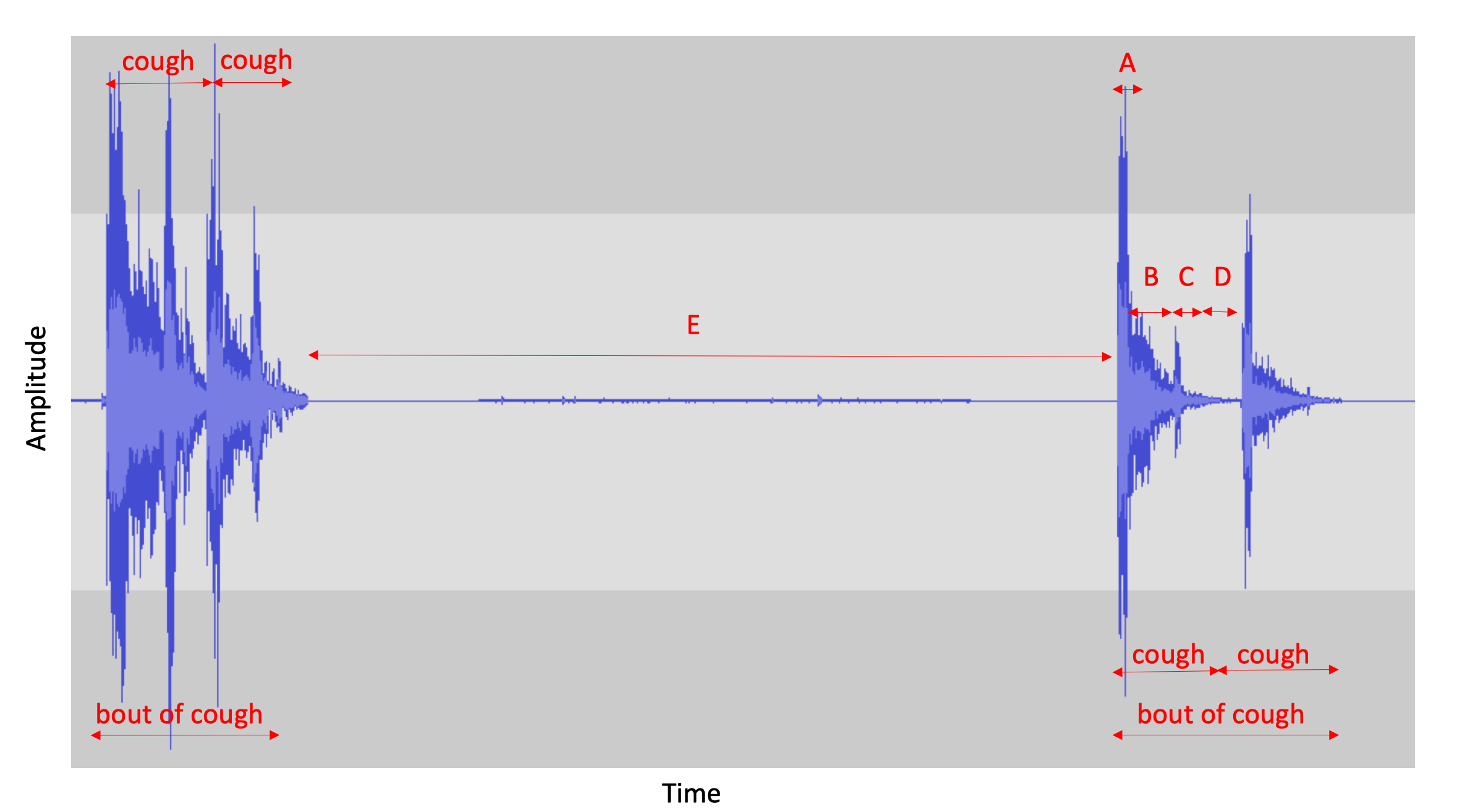}
\caption{Different states of coughing in an acoustic signal of four cough epochs} \label{fig:acoustic-states}
\end{figure}

We divided each audio signal into 25 millisecond time slots (bins) and extracted the following information from each bin: the time corresponding to the mid--point of each bin, the sum of the energy density of frequencies under 2 KHz (low-band energy), the sum of the energy density of frequencies between 2 KHz and 4 KHz (mid-band energy) and -- finally -- the sum of the energy density of frequencies between 4 KHz and 22 KHz (high-band energy). In light of the limited span of the audio frequency range, no frequencies above 22 KHz were considered. We considered the sum of energy densities as our training features for the Hidden Markov model. 
In this work, each cough recording was divided into five distinct states/stages, and each 25ms time bin was labeled as to the state with which it was associated.  Specifically, we considered three states inside a single cough (states A, B and C), a brief state of silence between each cough inside a bout of coughs (D) and a longer state of silence between bouts of coughs for cough-prone cases (E). Bouts of coughing were considered to trigger additional coughing (thus returning from state D to state A) with higher probability than in a general non-coughing state (state E); alternatively, a bout of coughing could then end, via a transition to state E.  Figure \ref{fig:acoustic-states} depicts different coughing states in the time domain.   B contrast, a schematic diagram showing posited transitions between different coughing states is demonstrated in Figure \ref {fig:hmm}. 

The length of the cough sounds vary from cough-to-cough, and the distinctions between the successive stages are not always clear -- leading to imprecision in human classification of such stages. The beginning of the cough sound was used as the starting point of state A, the start of state B was selected when the sound amplitude was significantly lower than the initial peak and the start of state C was chosen when there was a rise in the sound amplitude after state B.

This work sought to investigate the effectiveness of an HMM in predicting the underlying state of a given time interval of a cough--recording by feeding our model with low, mid and high band energy-density values. Given the characteristics of a single 25ms bin and the energy density values, we investigated the capacity of that model to predict with which state of coughing this bin was associated.

\subsection{Model Training}
The calculated probability for each hidden state is obtained by multiplying two values; one inferred from the observation i.e., the likelihood of observing that hidden state given the current observation vector and the other one derived from the transition matrix -- i.e., the probability of being in that specific state according to the probability of having been in different states in the previous time bin. 

\subsection{Model evaluation}
We employed a two-fold cross validation approach for training our model and and used AUC -- Area Under the Receiver Operating Characteristic [ROC] Curve -- as the primary evaluation metrics.  The confusion matrix, sensitivity, and specificity were considered to further evaluate model performance.

Since the ultimate goal is of this work to classify Cough from Non-Cough (correctly identifying an epoch of cough in a bout of coughs) or Coughing from Non-Coughing (correctly identifying a bout of coughs), we further investigated the capacity to classify audio signals according to two dichotomous categories:  Cough vs. Non-Cough, and Coughing vs. Non-Coughing.  To accomplish this, we grouped the states in binary format as follows:
\begin{itemize}
  \item \textbf{Cough vs. Non-Cough}: states A, B and C were grouped in a single state of Cough and states D and E as a single state of Non-Cough
  \item \textbf{Coughing vs. Non-Coughing}: States A, B, C and D were grouped as sate of Coughing and E as the state of Non-Coughing. 
\end{itemize}

The details of the preferred classifier will differ depending on our goals.  For example, one can either maximize sensitivity at the expense of specificity in order to have a model that is extremely effective at recognizing events identified as coughs (or coughing), but produces a lot of false positives. Likewise, the goal can be maximizing the specificity and obtaining a model that is subject to few false positives, but at the cost of large number of false negatives.
Here, we applied the Youden's index (Youden's J statistic) to maximize both sensitivity and specificity. The Confusion matrix and the optimal accuracy, sensitivity and specificity are demonstrated in table \ref{table:univariate}.

\subsubsection {Transition and Emission Matrices}
\begin{figure}
\includegraphics[width=0.6\linewidth,center]{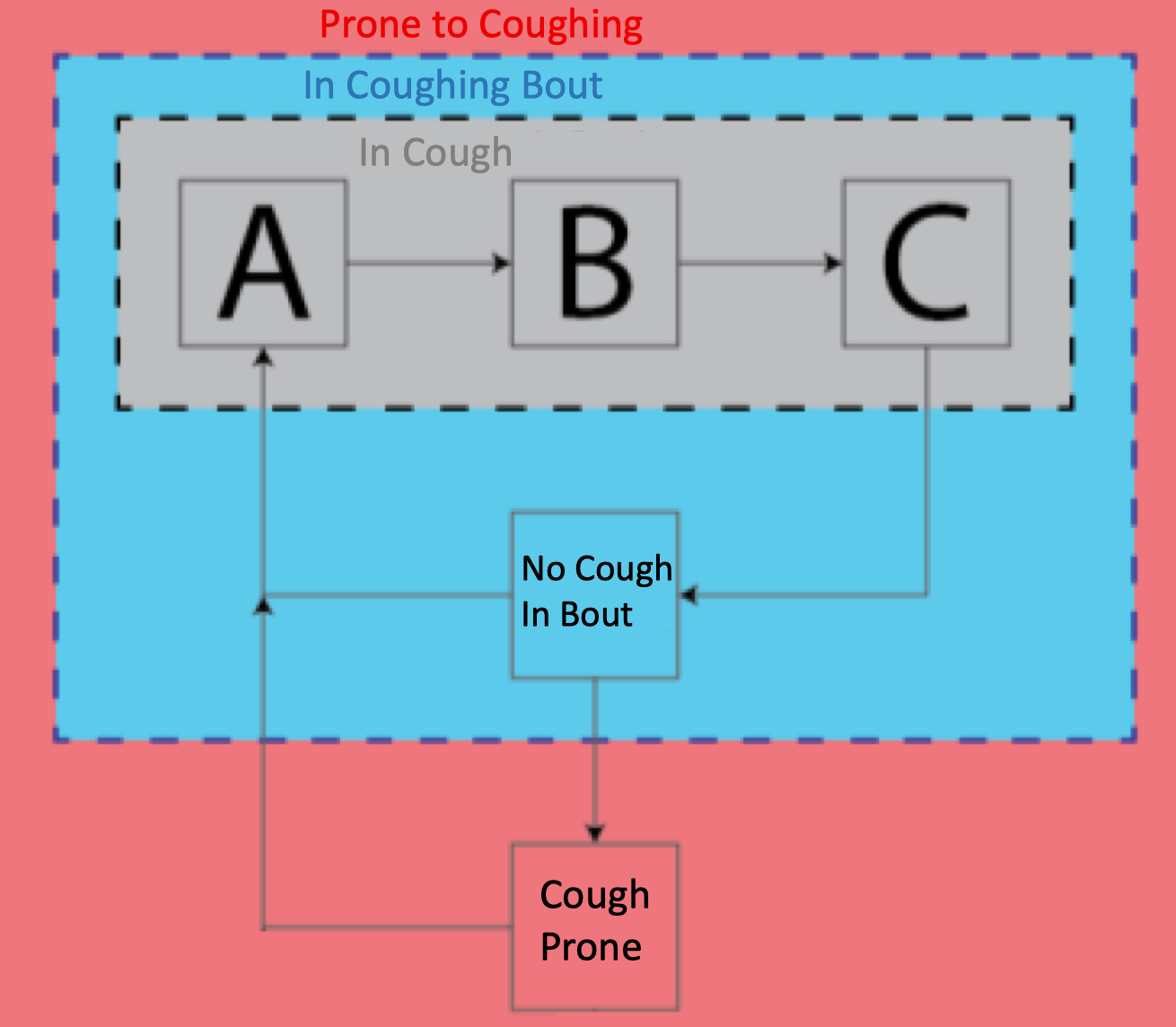}
\caption{Cough Transitions Captured in the Hidden Markov model} \label{fig:hmm}
\end{figure}

Table \ref{tab:train_data} shows a sample of data points extracted from cough signals. The HMM states and transitions captured the posited structure of transition ins between cough stages as shown in Figure \ref {fig:hmm}. 

\begin{table}[!ht]
\caption {Training Data Sample} \label{tab:train_data} 
    \begin{center}
        \begin{tabular}{|c|c|c|c|}
            \hline
            Ground Truth Label & Low--band energy & Mid--band energy & High--band energy \\
            \hline
            A & 31855.84538 & 1155.993164 & 678.3858756	\\
            B & 5630.510263	& 895.4698416 & 1704.088542 \\	
            B & 9672.263893	& 1891.187844 &	1126.828215	\\
            C & 371.2424149	& 8.467165454 & 2.065281857	\\
            D & 189.6220895	& 6.647450046 & 1.222387307	\\
            E & 3.117369203	& 0.389733216 & 0.058569281	\\
            E & 2.158550463 & 0.153855478 & 0.048496733 \\
            E & 1.134557944	& 0.095394591 &	0.020540539	\\
            \hline
        \end{tabular}
    \end{center}
\end{table}

At any given time bin, the HMM can be in one of the five (hidden) states of A, B, C, D or E, resulting in the transition matrix shown as table \ref{table:transition_matrix}. It bears emphasis that there are no transitions between some pairs of states -- for example, from A to C, or A to D); the probability of such transitions was treated as zero. 

\begin{table}[ht]
\caption {Transition Table for Sample Data} \label{table:transition_matrix} 
\begin{center}
\begin{tabular}{|c|c|c|c|c|c|}
\hline
& A & B & C & D & E\\
\hline
A & $P_{A|A}$ & $P_{A|B}$ & 0.0 & 0.0 & 0.0 \\
B & 0.0 & $P_{B|B}$ & $P_{B|C}$ & 0.0 & 0.0 \\
C & 0.0 & 0.0 & $P_{C|C}$ & $P_{C|D}$ & 0.0 \\
D & $P_{D|A}$ & 0.0 & 0.0 & $P_{D|D}$ & $P_{D|E}$\\
E & $P_{E|A}$ & 0.0 & 0.0 & 0.0 & $P_{E|E}$\\
\hline
\end{tabular}
\end{center}
\end{table}

To calculate the probability $P_{xy}$ of transition from a current state $x$ to any of the probable states $y$, we first found the probability of leaving a given state to any destination.  Based on the HMM assumption of memoryless transition processes, this is given by the reciprocal of the mean residence time (in time bins) within that state.  For states exhibiting a single outgoing transition (states A, B, C and E), that probability was employed directly.  For state D (which can be followed by either state A and state E), to arrive at the probability of making the transition to each of states A and E, we further multiplied the probability of leaving the state by the empirically observed proportion of transitions from state D to states A and E, respectively. 

Since the model in this work makes use of continuous observations, instead of having an emission matrix, we used density functions extracted from and fitted to empirical observations, where the observations are assumed to be independent from each other, conditional on being in a given state.  As a simplifying assumption, the joint likelihood of observing a given vector of low-band, mid-band and high-band energy quantities was approximated as the product of independent likelihood functions (each associated with a univariate probability density function). For a case of univariate HMM where a single observation (i.e, the total energy inside each bin), for any given state, only one empirical density function was defined.

\section{Results}
Two experiments were conducted using the HMM. Experiment A trained and evaluated a univariate HMM considering just a single feature: the total energy in a time-binned audio signal.  By contrast, in Experiment B, all the three band of energies were considered as a vector of observations, and a multivariate HMM was trained.  Both experiments used the ``mhsmm'' package in the statistical software R.  Both Experiments evaluated the HMMs according to ability to classify, for a given time bin, the particular coughing state as well as dichotomous classification regarding the presence of absence of coughing.

\subsection{Results of the univariate HMM:  Experiment A }\label{subsec:univariate_HMM}
\begin{figure}%
\centering
\subfloat[Training Set]{{\includegraphics[width=.45\linewidth]{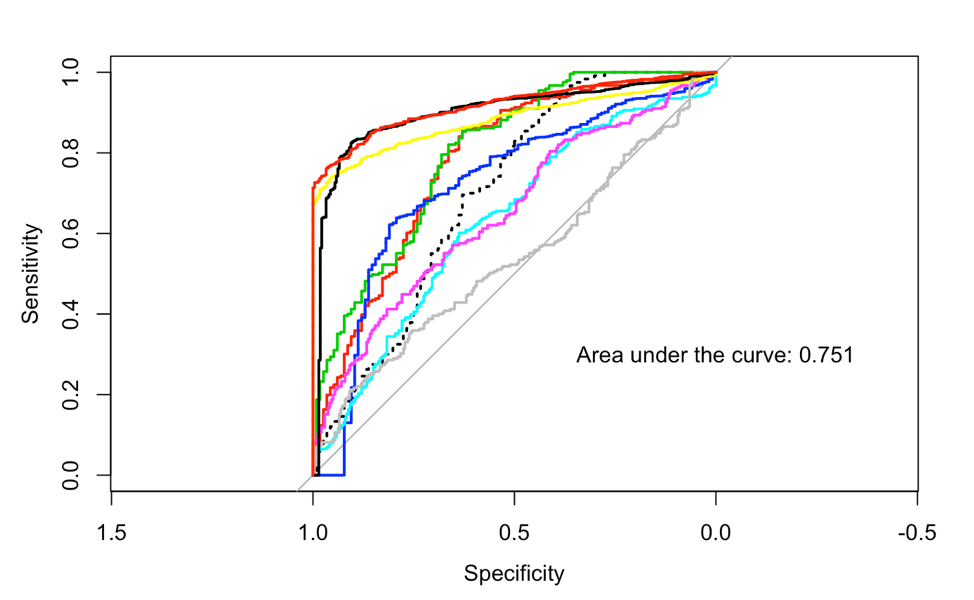} }}%
\qquad
\subfloat[Testing Set]{{\includegraphics[width=.45\linewidth]{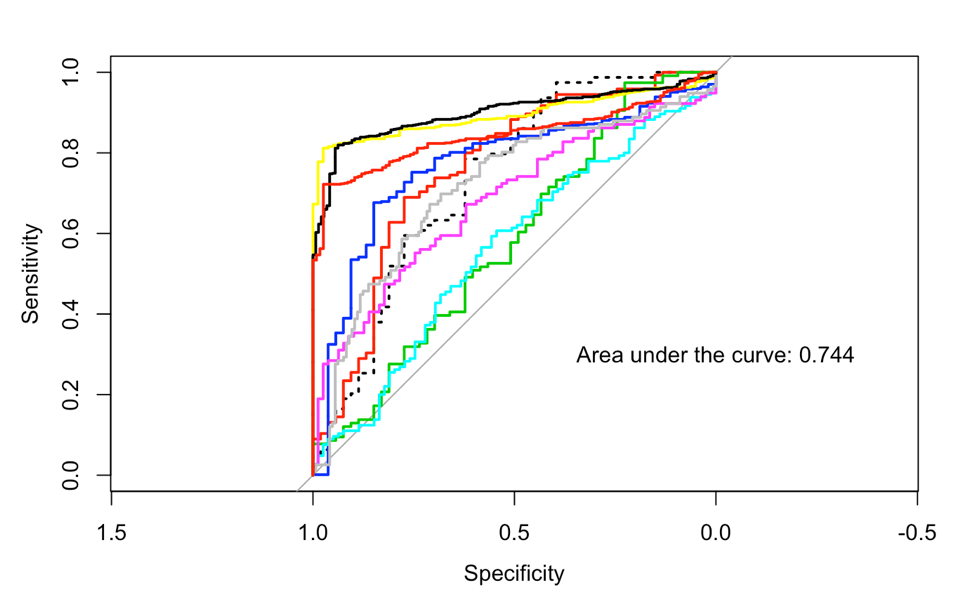} }}%
\caption{multi-class ROC curves in one-vs-one fashion for a univariate HMM}%
\label{fig:multiROC_uni_HMM}%
\end{figure}

Using the total energy in bins as the single feature, an AUC value of 0.751 and 0.744 was obtained for training and testing sets, respectively. The performance statistics of the model over the testing set -- including a confusion matrix, sensitivity, specificity, and accuracy -- is shown in table~\ref{tab:confusion_matrix_sensi_univariate_testing} and the multiclass ROC curves in a one-vs-one class fashion for training and testing sets are demonstrated in figure~\ref{fig:multiROC_uni_HMM}. 

\begin{table}[ht]
\caption {Performance statistics of the testing set for univariate HMM} \label{tab:confusion_matrix_sensi_univariate_testing} 
\begin{center}
\begin{tabular}{|c|c|c|c|c|c|c|}
\hline
\multicolumn{2}{|c|}{} & \multicolumn{5}{|c|}{Observed}\\
\hline
\multicolumn{2}{|c|}{} & Class: A & Class: B & Class: C & Class: D & Class: E\\
\hline
\multirow{ 5}{*}{Predicted} & Class: A & 31 & 6 & 1 & 1 & 7\\
& Class: B & 3 & 45 & 19 & 6 & 25 \\
& Class: C & 3 & 17 & 29 & 4 & 5\\
& Class: D & 3 & 2 & 31 & 21 & 19\\
& Class: E & 13 & 9 & 65 & 84 & 714\\
\hline
\hline
\multicolumn{2}{|c|}{Sensitivity} & 0.58491 & 0.56962 & 0.20000 & 0.18103 & 0.9273\\
\hline
\multicolumn{2}{|c|}{Specificity} & 0.98649 & 0.95111 & 0.97151 & 0.94747 & 0.5649 \\
\hline
\multicolumn{2}{|c|}{Accuracy} & \multicolumn{5}{|c|}{0.7223}\\
\hline
\end{tabular}
\end{center}
\end{table}


\begin{figure}%
\centering
\subfloat[Cough/Non\_cough; AUC 0.844]{{\includegraphics[width=.45\linewidth]{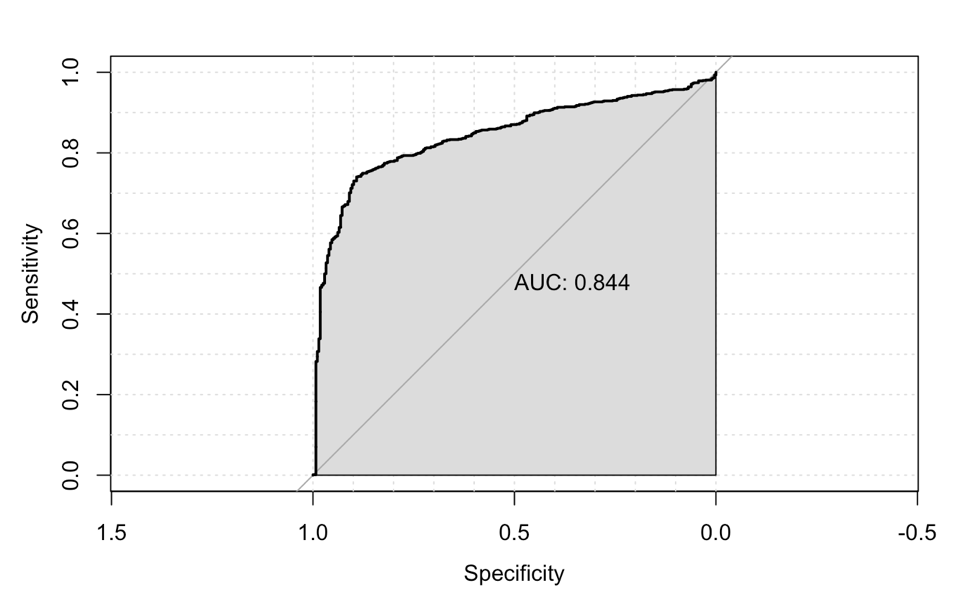}}}%
\qquad
\subfloat[Coughing/Non\_coughing; AUC 0.865]{{\includegraphics[width=.45\linewidth]{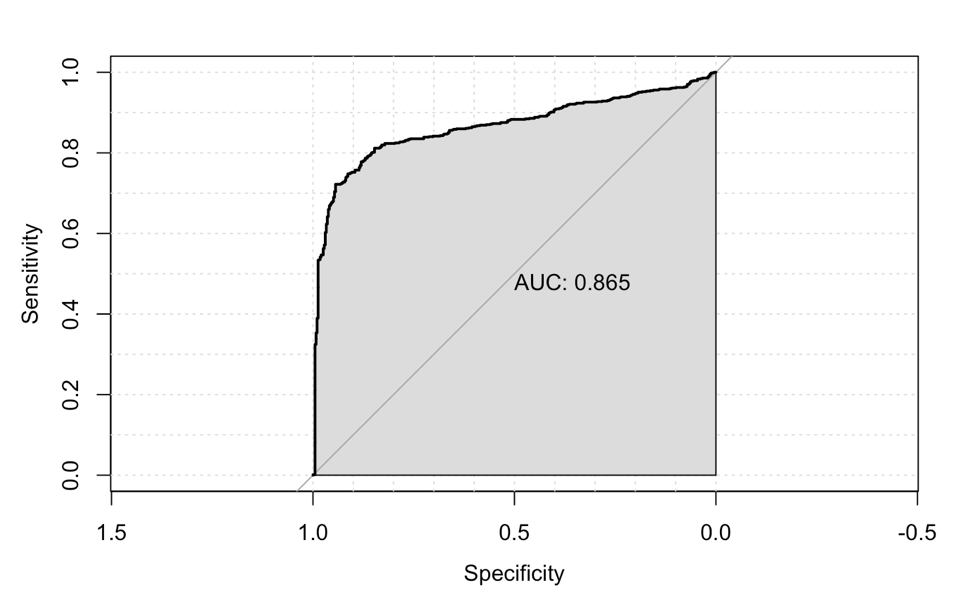}}}%
\caption{ROC curve for uni-variate HMM after grouping}%
\label{fig:ROC_uni_HMM}%
\end{figure}

Such a grouping process resulted in the following ROC curves shown in figure~\ref{fig:ROC_uni_HMM} for Cough/Non-Cough and Coughing/Non-Coughing classifications. 

\begin{table}
\caption{Performance statistics of the testing set for the univariate HMM in cough/no\_cough and coughing/no\_coughing classification mode}
\label{table:univariate}
\begin{tabular}{|p{3.5cm}|>{\centering}p{1cm}p{2cm}p{2cm}>{\centering}p{1.4cm}|>{\centering}p{1cm}p{0.5cm}p{0.5cm}p{1.9cm}|}

\cline{1-9}
    \multirow{3}{*}{} & &\multicolumn{2}{C{4cm}}{Identifying a cough epoch in bout of coughs} & & &\multicolumn{2}{C{2.5cm}}{Identifying a bout of coughs} &\\
\cline{2-9} 
      &\multicolumn{3}{c}{Observed}& & &\multicolumn{2}{c}{Observed} &\\
\cline{2-9} 
    & cough & \multicolumn{2}{c}{} & no--cough &  coughing &\multicolumn{2}{c}{}& no--coughing\\
\hline
    Predicted cough(ing) & 247 & \multicolumn{2}{c}{} & 230 & 371 & \multicolumn{2}{c}{} & 214 \\
    Predicted no-cough(ing) & 30 & \multicolumn{2}{c}{} & 656 & 22 & \multicolumn{2}{c}{} & 556 \\
    & &\multicolumn{2}{c}{Accuracy: $78\%$} & & & \multicolumn{2}{c}{Accuracy: $80\%$} & \\
    & &\multicolumn{2}{c}{Sensitivity: $89\%$} & & & \multicolumn{2}{c}{Sensitivity: $94\%$} &\\
     & & \multicolumn{2}{c}{Specificity: $74\%$} & & &  \multicolumn{2}{c}{Specificity: $72\%$} &\\
\hline
\end{tabular}
\end{table}



\subsection{Multivariate HMM Results:  Experiment B}
The multivariate HMM trained with a vector of three features containing the acoustic energy in low, medium and high bands improved by 6\% the performance of the AUC for the testing set, increasing it from 0.744 to 0.789. The AUC for training set was almost the same as for the univariate case, reaching 0.752. The performance statistics of the chosen by Youden's-index-selected multivariate model over the testing set is demonstrated at table~\ref{tab:confusion_matrix_sensi_multivariate_testing}. 

\begin{table}[ht]
\caption {Performance statistics of the testing set for multivariate HMM} \label{tab:confusion_matrix_sensi_multivariate_testing}
\begin{center}
\begin{tabular}{|c|c|c|c|c|c|c|}
\hline
\multicolumn{2}{|c|}{} & \multicolumn{5}{|c|}{Observed}\\
\hline
\multicolumn{2}{|c|}{} & Class: A & Class: B & Class: C & Class: D & Class: E\\
\hline
\multirow{ 5}{*}{Predicted} & Class: A & 41 & 12 & 0 & 1 & 5\\
& Class: B & 4 & 41 & 6 & 0 & 8 \\
& Class: C & 0 & 21 & 33 & 2 & 10\\
& Class: D & 2 & 4 & 43 & 26 & 3\\
& Class: E & 6 & 1 & 63 & 87 & 744\\
\hline
\hline
\multicolumn{2}{|c|}{Sensitivity} & 0.77358 & 0.51899 & 0.22759 & 0.22414 & 0.9662\\
\hline
\multicolumn{2}{|c|}{Specificity} & 0.98378 & 0.98339 & 0.96758 & 0.95033 & 0.6005 \\
\hline
\multicolumn{2}{|c|}{Accuracy} & \multicolumn{5}{|c|}{0.761}\\
\hline
\end{tabular}
\end{center}
\end{table}


\begin{figure}%
\centering
\subfloat[Training Set]{{\includegraphics[width=.45\linewidth]{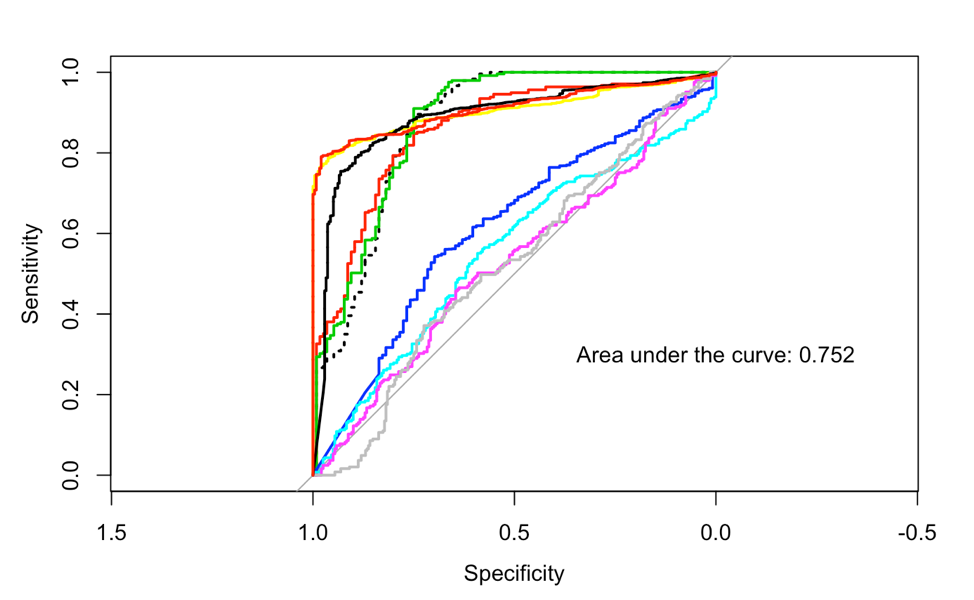} }}%
\qquad
\subfloat[Testing Set]{{\includegraphics[width=.45\linewidth]{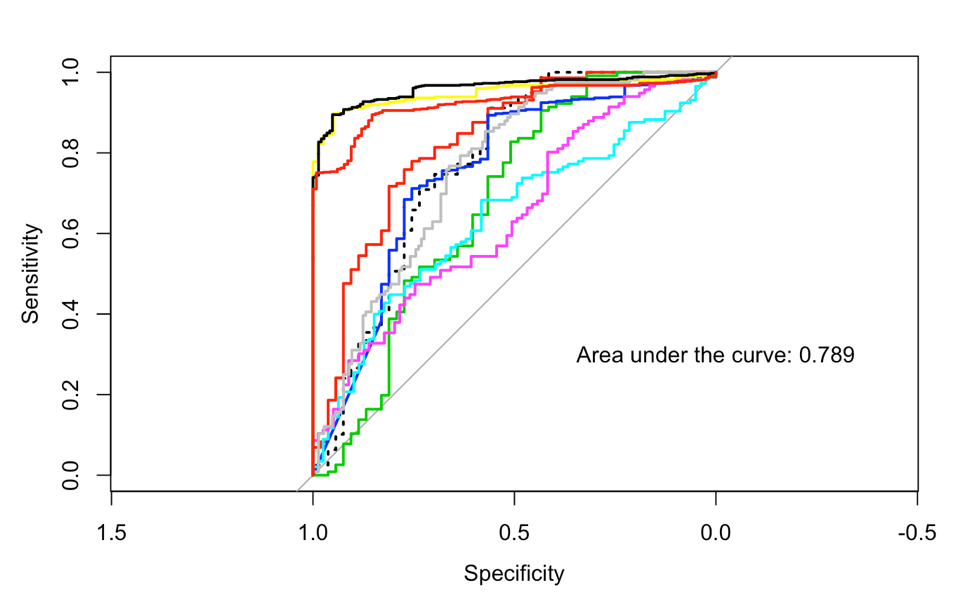} }}%
\caption{multi-class ROC curves in one-vs-one fashion for a multivariate HMM}%
\label{fig:multiROC_multi_HMM}%
\end{figure}

\begin{figure}%
\centering
\subfloat[Cough/Non\_cough; AUC 0.882]{{\includegraphics[width=.45\linewidth]{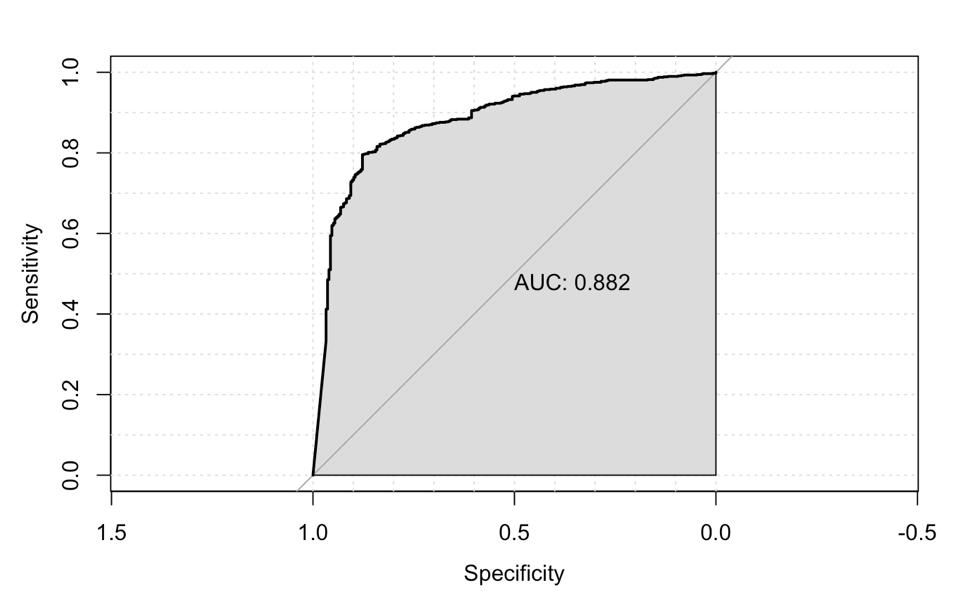}}}%
\qquad
\subfloat[Coughing/Non\_coughing; AUC 0.920]{{\includegraphics[width=.45\linewidth]{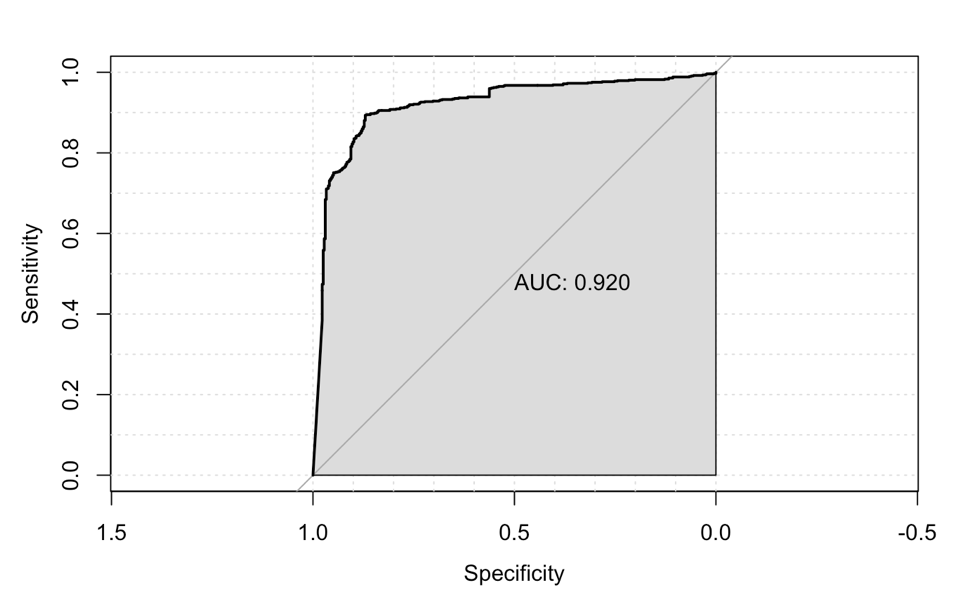}}}%
\caption{ROC curve for multivariate HMM after grouping}%
\label{fig:ROC_multi_HMM}%
\end{figure}

Again to investigate the obtained models performance in classifying Cough from Non-Cough or Coughing from Non-Coughing, the identified states were grouped as per the process discussed in Section ~\ref{subsec:univariate_HMM}.
Figure~\ref{fig:ROC_multi_HMM} shows the results of the Cough/Non-Cough and Coughing/Non-Coughing classifications as the results of dichotomously grouping the cough states. The AUC for the cases of Cough/Non-Cough and Coughing/Non-Coughing classification were increased by 4.5\% and 6.4\% when compared to their univariate HMM counterparts. 

Using the curves demonstrated in figure~\ref{fig:ROC_multi_HMM} and to maximize both the sensitivity and specificity, the best cut-off point was calculated on which the confusion matrix and the optimal accuracy, sensitivity and specificity were obtained, according to Youden's index. Results of using the best threshold in terms of balancing the sensitivity and specificity are shown in table \ref{table:multivariate}.

\begin{table}
\caption{Performance statistics of the testing set for multi-variate HMM in cough/no\_cough and coughing/no\_coughing classification mode}
\label{table:multivariate}
\begin{tabular}{|p{3.5cm}|>{\centering}p{1cm}p{2cm}p{2cm}>{\centering}p{1.4cm}|>{\centering}p{1cm}p{0.5cm}p{0.5cm}p{1.9cm}|}

\cline{1-9}
    \multirow{3}{*}{} & &\multicolumn{2}{C{4cm}}{Identifying a cough epoch in bout of coughs} & & &\multicolumn{2}{C{2.5cm}}{Identifying a bout of coughs} &\\
\cline{2-9} 
      &\multicolumn{3}{c}{Observed}& & &\multicolumn{2}{c}{Observed} &\\
\cline{2-9} 
    & cough & \multicolumn{2}{c}{} & no--cough &  coughing &\multicolumn{2}{c}{}& no--coughing\\
\hline
    Predicted cough(ing) & 243 & \multicolumn{2}{c}{} & 181 & 342 & \multicolumn{2}{c}{} & 82 \\
    Predicted no-cough(ing) & 34 & \multicolumn{2}{c}{} & 705 & 51 & \multicolumn{2}{c}{} & 668 \\
    & &\multicolumn{2}{c}{Accuracy: $82\%$} & & & \multicolumn{2}{c}{Accuracy: $89\%$} & \\
    & &\multicolumn{2}{c}{Sensitivity: $88\%$} & & & \multicolumn{2}{c}{Sensitivity: $87\%$} &\\
     & & \multicolumn{2}{c}{Specificity: $80\%$} & & &  \multicolumn{2}{c}{Specificity: $90\%$} &\\
\hline
\end{tabular}
\end{table}

\section{Conclusion}
The HMMs evaluated here demonstrated favorable results, especially when the obtained results were interpreted as a dichotomously problem of distinguishing Coughs from Non\_Coughs, or  Coughing from Non\_Coughing periods. Moreover, the multivariate HMM performed slightly more favourably than did a univariate HMM.

Unsurprisingly, the results presented in this work further suggest that the multivariate HMM demonstrates classification and detection of cough events with higher accuracy than does a to univariate HMM. Splitting the energy of cough sounds into three separate bands lead to density functions corresponding to each band which can provide more detailed information to the HMM.  

While the results presented here demonstrate much promise, the approach applied exhibits significant limitations and room for improvements.  The added accuracy associated with multivariate analysis invites investigation into both alternative bands, but also classification according to a larger number of such bands.  The library of cough sounds examined here were greatly limited in their sourcing; results presented here may differ significantly for alternative coughing etiologies, and according to the pulmonary and upper-respiratory character and physical shape of the individual coughing, and potentially according to cultural norms involved.  Greater variety in sourcing of cough source remains a high priority.  Moreover, the classification accuracy exhibited in this study needs to be considered in light of the limited  library of recordings employed here; other audio recordings containing a variety of background noise or other respiratory-related sounds may exhibit marked difference in the accuracy of classification that they support using similar HMMs.  Finally, it will be important to consider examining other classifiers, that provide additional avenues for predictive accuracy, including classifiers that are less theory-based, such as artificial recurrent neural networks or deep learning networks employing recurrent network structures.

Despite its limitations, the cough analysis presented approach can provide a foundation towards support both clinical research on pulmonary distress at a clinical level and for capturing patient outcomes.  It further offers intriguing potential for early-warning outbreak detection in public areas using mobile sensor data -- such as from wearable devices and smartphones, particularly when coupled with transmission modeling and tools such as particle filtering. Another potential application of this study can be symptomatically-triggered treatment of patients suffering from respiratory diseases, particularly in patients that lack ready capacity to communicate their distress, such as in infants and young children, and among adults suffering from dementia or verbal limitations.  The technique also offers potential for recognizing animal vocalization and diagnosing animal health status.

%
%
%
%

\end{document}